\documentclass[twocolumn,superscriptaddress,prl,showpacs,float]{revtex4-2}
\usepackage{amsmath}
\usepackage{amssymb}
\usepackage{graphicx}
\usepackage{hyperref}
\usepackage{color}

\begin{document}

\title{Magnetic-field-controlled spin fluctuations and quantum criticality in Sr$_3$Ru$_2$O$_7$}

\author{C. Lester}
\affiliation{H.H. Wills Physics Laboratory, University of Bristol, Tyndall Ave., Bristol, BS8 1TL, UK}

\author{S. Ramos}
\affiliation{School of Physical Sciences, University of Kent, Canterbury, CT2 7NH, UK}

\author{R. S. Perry}
\affiliation{London Centre for Nanotechnology, University College London, London, WC1H 0AH, UK}

\author{T. P. Croft}
\affiliation{H.H. Wills Physics Laboratory, University of Bristol, Tyndall Ave., Bristol, BS8 1TL, UK}

\author{M. Laver}
\affiliation{School of Physics and Astronomy, University of Birmingham, Birmingham B15 2TT, UK}

\author{R. I. Bewley}
\affiliation{ISIS Facility, Rutherford Appleton Laboratory, Chilton, Didcot, OX11 0QX, UK}

\author{T. Guidi}
\affiliation{ISIS Facility, Rutherford Appleton Laboratory, Chilton, Didcot, OX11 0QX, UK}

\author{A. Hiess}
\altaffiliation{Current address: European Spallation Source ERIC, P.O. Box 176, 22100 Lund, Sweden}
\affiliation{Institut Laue-Langevin, 71 avenue des Martyrs, CS 20156, 38042 Grenoble, France}

\author{A.  Wildes}
\affiliation{Institut Laue-Langevin, 71 avenue des Martyrs, CS 20156, 38042 Grenoble, France}

\author{E. M. Forgan}
\affiliation{School of Physics and Astronomy, University of Birmingham, Birmingham B15 2TT, UK}

\author{S. M. Hayden}
\email{s.hayden@bristol.ac.uk} 
\affiliation{H.H. Wills Physics Laboratory, University of Bristol, Tyndall Ave., Bristol, BS8 1TL, UK}


\begin{abstract}
When the transition temperature of a continuous phase transition is tuned to absolute zero, new ordered phases and physical behaviour emerge in the vicinity of the resulting quantum critical point.
Sr$_3$Ru$_2$O$_7$ can be tuned through quantum criticality with magnetic field at low temperature. Near its critical field $B_c$ it displays the hallmark $T$-linear resistivity and a $T \log(1/T)$ electronic heat capacity behaviour of strange metals. However, these behaviours have not been related to any critical fluctuations. Here we use inelastic neutron scattering to reveal the presence of collective spin fluctuations whose relaxation time and strength show a nearly singular variation with magnetic field as $B_c$ is approached. The large increase in the electronic heat capacity and entropy near $B_c$ can be understood quantitatively in terms of the scattering of conduction electrons by these spin-fluctuations.  On entering the spin-density-wave ordered phase present near $B_c$, the fluctuations become stronger suggesting that the order is stabilised through an ``order-by-disorder'' mechanism.
\end{abstract}
\maketitle

The nature of the quantum criticality in Sr$_3$Ru$_2$O$_7$ has been debated for twenty years, since the discovery of its anomalous field-dependent electronic properties \cite{Grigera2001_GPSC,Grigera2004_GGBW,Rost2009_RPMM,Borzi2007_BGFP}. Early proposals suggested that critical fluctuations were associated with metamagnetism \cite{Grigera2001_GPSC} or fluctuations in the Fermi surface \cite{Grigera2004_GGBW}.  However, the discovery of spin density wave (SDW) order in Sr$_3$Ru$_2$O$_7$ \cite{Lester2015_LRPC} brought these proposals into question.  

The application of a large magnetic field to metals that have antiferromagnetic or spin-density wave order typically destabilises this order because the magnetic moments tend to align with the field. However certain metals, including Sr$_3$Ru$_2$O$_7$ \cite{Lester2015_LRPC} and URu$_2$Si$_2$ \cite{Knafo2016_KDBK}, have recently been discovered to exhibit magnetic order that is favoured by the magnetic field over a small range of field values.  In the case of Sr$_3$Ru$_2$O$_7$, this tuning effect of the field is believed to arise because of a field-induced Lifshitz transition in the Fermi surface \cite{Efremov2019_ESRC} which changes the nesting and the wavevector-dependent susceptibility $\chi(\mathbf{Q})$. Spin orbit coupling is also important as shown by the strong dependence of the SDW order on the magnetic field direction \cite{Lester2015_LRPC}.

\begin{figure*}[t]
\begin{center}
\includegraphics[width=0.99\linewidth]{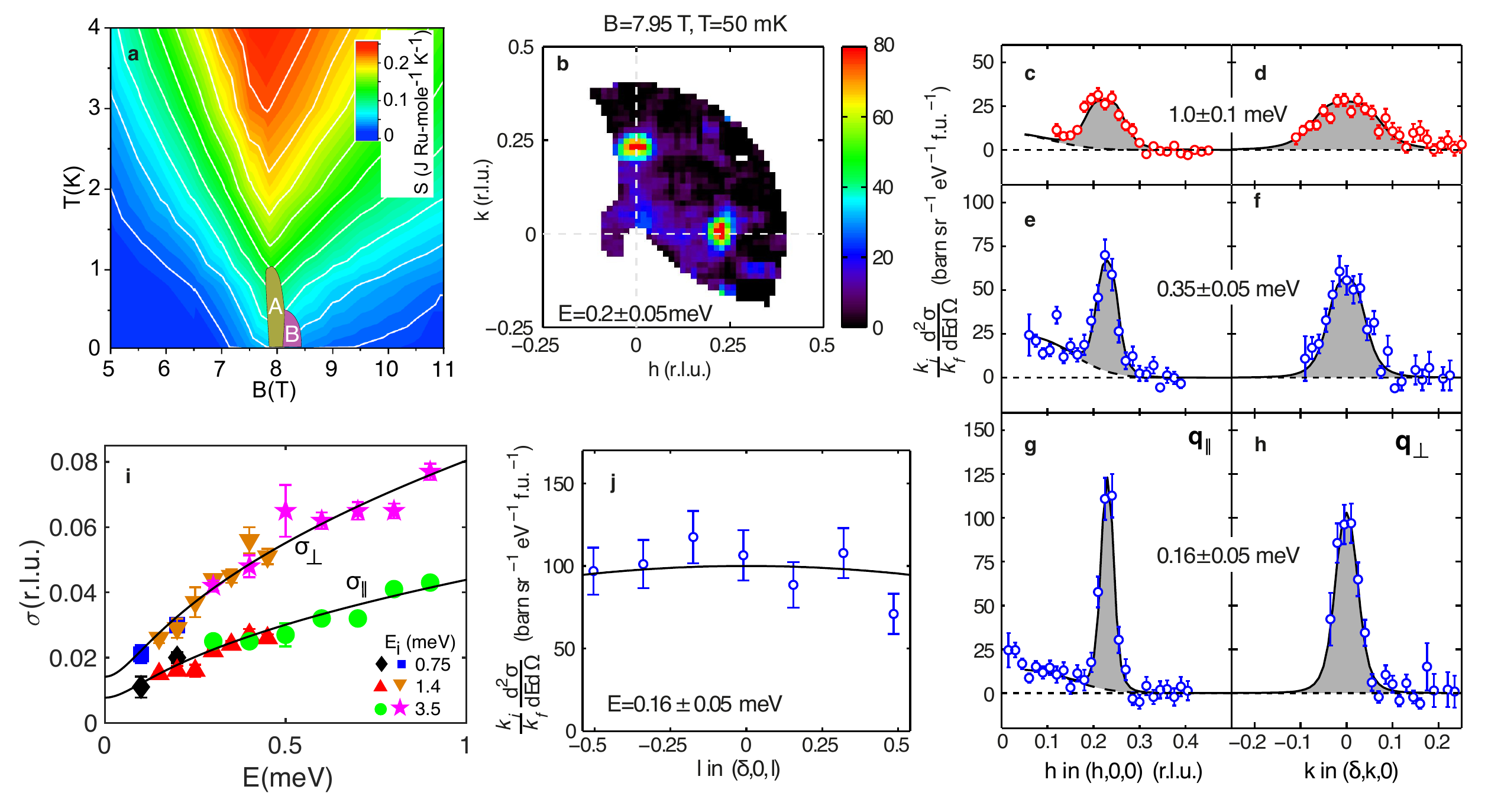}
\end{center}
\caption{\textbf{Quantum criticality and the wavevector-dependence of the magnetic excitations}. \textbf{a} Field and temperature dependence of entropy (based on data of Sun et al. \cite{Sun2018_SRPM}) showing a quantum critical behaviour near $B_c \approx 7.95$~T and SDW phases A and B \cite{Grigera2004_GGBW, Lester2015_LRPC}. \textbf{b} Wavevector-dependence of low-energy (0.2~meV) spin fluctuations measured at $B_c$ showing strong excitations near $\mathbf{Q}_{\delta}$ (units as c-h). \textbf{c}-\textbf{h}  $\mathbf{Q}$-cuts through $\mathbf{Q}_{\delta}$ for different energies showing the dispersion of the excitations. The direction of the cut with respect to $\mathbf{Q}_{\delta}$ is denoted by $\|$ and $\perp$. \textbf{i} Energy dependence of the half-width-half-maximum $\sigma$ of peaks such as those in \textbf{c}-\textbf{h}. Solid line is a fit to the phenomenological MMP model susceptibility described in the text with the parameters in Table.~\ref{table:fitted_parms}. \textbf{j} The 2D nature of the spin fluctuations is demonstrated by the lack of $\ell$-dependence of the response. Data in panels \textbf{c}-\textbf{h} have been integrated over the range $-0.5<\ell<0.5$.  In this paper, we label reciprocal space $(h,k,\ell)$ in units of $(2 \pi/a,2 \pi /b,2 \pi /c)$ using the $I4/mmm$ space group with \cite{Shaked2000_SJCI} $a \approx 3.89$~\AA\ and $c \approx 20.7$~\AA. Error bars are determined from Poisson counting statistics or least squares fitting of data and denote one standard deviation.}
\label{FIG:Qcuts_with_AB_phase}
\end{figure*}

When a magnetic field is applied along the $c$-axis,  that is perpendicular to the RuO$_2$ planes, two SDW phases (A and B) are observed \cite{Lester2015_LRPC} for $T \lesssim $ 1\;K and $7.8 \lesssim B \lesssim 8.5$~T. The $B-T$ phase diagram is shown in Fig.~\ref{FIG:Qcuts_with_AB_phase}a.  Transport and thermodynamic measurements \cite{Grigera2001_GPSC,Tokiwa2016_TMPG, Sun2018_SRPM} indicate a large region of temperature-induced fluctuations surrounding the SDW ordered phases. 
There is a quantum critical fan centred on $B_c \sim 7.95$\;T as shown in the field and temperature dependent entropy \cite{Tokiwa2016_TMPG,Sun2018_SRPM} plotted in Fig.~\ref{FIG:Qcuts_with_AB_phase}a. Note the ridge at $B_c$ characteristic of quantum criticality. 
As $B \rightarrow B_c$, the low-temperature coefficient of specific heat \cite{Rost2009_RPMM} diverges as $\gamma=C/T \sim 1/|B-B_c|$ and the exponent \cite{Grigera2001_GPSC} of the resistivity $\alpha$ in $\rho=\rho_0+A T^{\alpha}$ falls to $\alpha=1$. At $B=B_c$ a low temperature divergence $C \sim T\log (1/T)$ is observed until it is cut off at the SDW ordering temperature \cite{Rost2011_RGBP}.    

In this work we investigate the very low-energy collective magnetic excitations to determine how they relate to the strange metal and quantum critical behaviour. We find that they are dominated by an overdamped transverse mode with in-plane fluctuations whose characteristic relaxation rate evolves in a nearly singular way. We use a phenomenological spin fluctuation model to relate this behaviour to the low-temperature thermal properties.

\section*{Results}

We study the very-low energy collective spin fluctuations of Sr$_3$Ru$_2$O$_7$ near the SDW ordering wavevector $\mathbf{Q}_{\delta}$ using inelastic neutron scattering (INS). Measurements are made over the energy range 0--1\;meV as a function of $B$.  We used the LET time-of-flight spectrometer at the ISIS neutron spallation source (see Methods). Fig.~\ref{FIG:Qcuts_with_AB_phase}b shows a \textbf{Q}-map of the magnetic scattering at $\hbar \omega=0.2$~meV for $B=B_c=7.95$~T and $T=50$~mK. Strong inelastic scattering can be seen near the SDW ordering wavevectors \cite{Lester2015_LRPC} $\mathbf{Q}_{\delta} =(\pm\delta,0,0)$ and $(0,\pm\delta,0)$ with $\delta=0.23$.  Higher-energy ($\gtrsim 1$\,meV) excitations with similar wavevectors have been previously observed in Sr$_3$Ru$_2$O$_7$ at $T=1.5$\,K \cite{Capogna2003_CFHW, Ramos2008_RFBH}.  Fig.~\ref{FIG:Qcuts_with_AB_phase}c-h show cuts through $\mathbf{Q}_{\delta}$ for various energies. It can be seen that the magnetic excitations broaden with energy. We can define the energy-dependent peak widths (half-width at half maximum) for cuts parallel and perpendicular to $\mathbf{Q}_{\delta}$ as $\sigma_{\|}(\omega)$ and $\sigma_{\perp}(\omega)$ respectively. These are determined from data such as that in Fig.~\ref{FIG:Qcuts_with_AB_phase}c-h and shown in  Fig.~\ref{FIG:Qcuts_with_AB_phase}i.  Fig.~\ref{FIG:Qcuts_with_AB_phase}j shows a cut along $\ell$ through $\mathbf{Q}_{\delta}$, the almost constant intensity indicates a lack of $\ell$-dependence of the response and is consistent with the moments in the RuO$_2$ bilayers fluctuating independently \cite{Capogna2003_CFHW} as previously observed at $B=0$ and higher energies. In our analysis we treat the system as being 2D and ignore the $\ell$ dependence of the excitations.

\begin{figure*}[t]
\begin{center}
\includegraphics[width=0.8\linewidth]{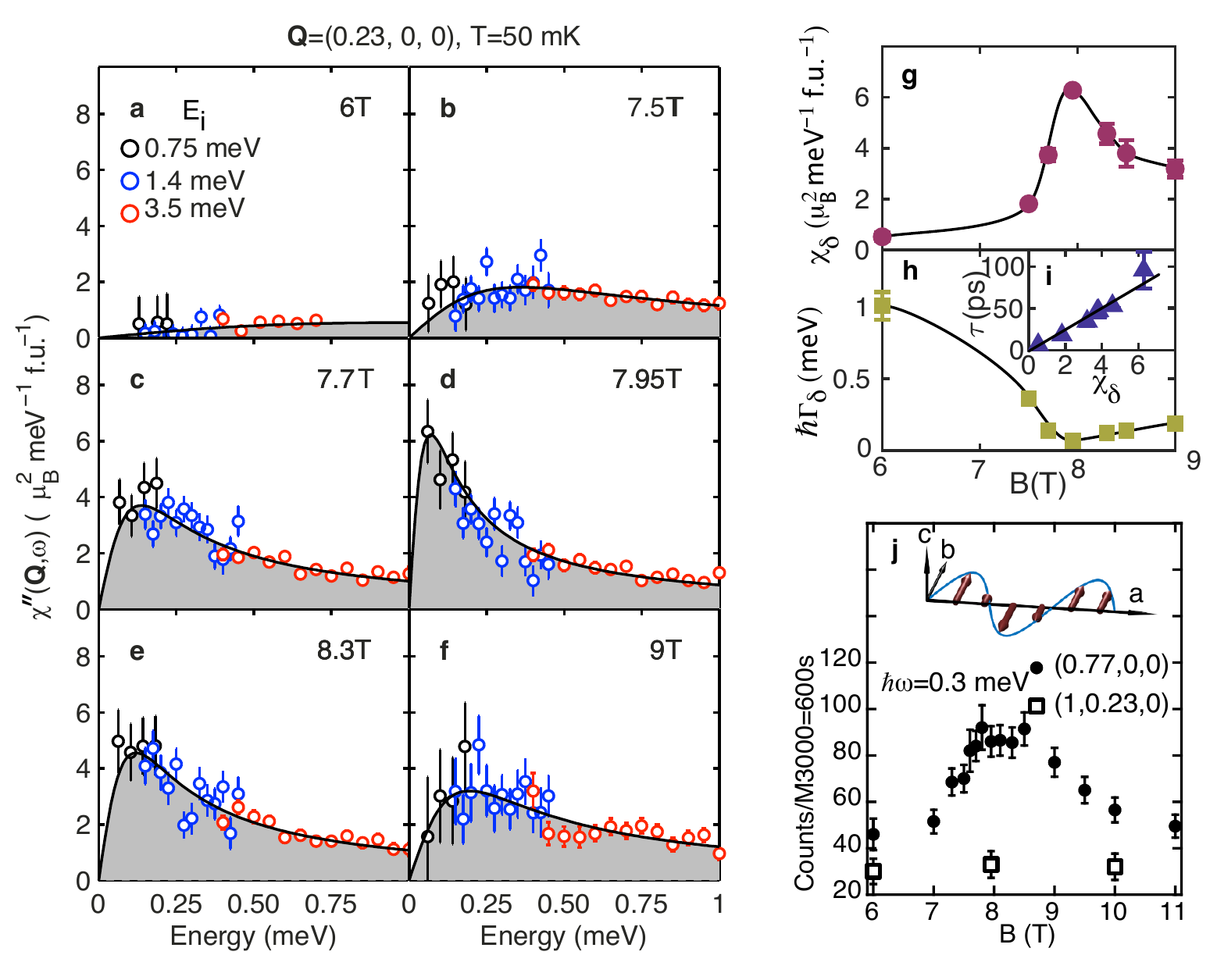}
\end{center}
\caption{\textbf{The collective spin fluctuations in Sr$_3$Ru$_2$O$_7$ measured through its quantum critical point.} \textbf{a-f} Energy-dependent spectra of the magnetic response function $\chi^{\prime\prime}(\mathbf{Q},\omega)$ for $\mathbf{Q}=\mathbf{Q}_\delta$  measured for magnetic fields through the critical field. The magnetic response has a Lorentzian form (solid lines) and softens (strengthens at the lowest energies) for the critical field $B_c=7.95$~T. The data have been integrated over $-0.5< \ell <0.5$. Colour of points indicates the incident neutron energy ($E_i$), also in Fig.\,\ref{FIG:Qcuts_with_AB_phase}. \textbf{g} The wavevector-dependent susceptibility $\chi_{\delta}$ is peaked at $B_c$. \textbf{h} The spin relaxation rate $\Gamma_{\delta}$ is minimum at $B_c$ and the inset shows that the relaxation time $\tau=1/\Gamma_{\delta} \propto \chi_{\delta}$. \textbf{j} The spin fluctuations are seen at $\mathbf{Q}=(1-\delta,0,0)$ and not at $(1,\delta,0)$ implying that they have transverse polarisation as illustrated for the SDW propagating along the $a$-axis.}
\label{FIG:Ecuts}
\end{figure*}

Data such as that in Fig.~\ref{FIG:Qcuts_with_AB_phase}b-h can be converted to the magnetic response function $\chi^{\prime\prime}(\mathbf{Q},\omega)$ (see Methods). Fig.~\ref{FIG:Ecuts}a-f shows the energy-dependence of $\chi^{\prime\prime}(\mathbf{Q},\omega)$ at $\mathbf{Q}_\delta$ for $B$ increasing through $B_c$. Our data is consistent with an overdamped (relaxational) magnetic response at all fields investigated which can be described by a Lorentzian energy dependence.  We observe a dramatic reduction in the energy scale of the magnetic response and an increase in its amplitude as the applied field approaches $B_c$.  This is direct evidence that the scattering we observe is associated with the quantum critical point at $B_c$. 

The observed response is characteristic of nearly antiferromagnetic metals (See Methods). It can be well described using a 2D phenomenological model used by Moriya \cite{Moriya1970_M,Moriya2000_MU} and Millis-Monien-Pines (MMP) \cite{Millis1990_MMP}, for convenience we call this the MMP form:
\begin{align}
      \chi^{\prime\prime}(\mathbf{Q},\omega) &=\sum_{\mathbf{Q}_{\delta}}\frac{\chi_{\delta} \Gamma_{\delta} \omega }
    { \Gamma_{\delta}^2  (1+\xi^2_{\|}q^2_{\|}+\xi^2_{\perp}q^2_{\perp})^2+\omega^2}  \label{EQN:MMP}. 
\end{align}
The MMP form has been widely used to describe cuprate superconductors \cite{Millis1990_MMP}. In the present case of incommensurate spin density fluctuations we have $\mathbf{Q}_{\delta}=(\pm \delta,0)$ and $(0, \pm \delta)$, rather than $(1/2 \pm \delta,1/2)$ and $(1/2, 1/2 \pm \delta)$ in the cuprates. The components of the reduced  wavevector $\mathbf{q}=\mathbf{Q}-\mathbf{Q}_{\boldsymbol{\delta}}$ parallel and perpendicular to $\mathbf{Q}_{\boldsymbol{\delta}}$ are $q_{\|}$ and $q_{\perp}$ respectively. The corresponding correlation lengths are $\xi_{\|}$ and $\xi_{\perp}$ respectively. Near $\mathbf{Q}_{\delta}$, Eqn.~\ref{EQN:MMP} yields excitations described by an overdamped harmonic oscillator with a relaxation rate $\Gamma(\mathbf{Q})$ given by
\begin{align}
\Gamma(\mathbf{Q})&=\Gamma_{\delta} \left(1+\xi^2_{\|}q^2_{\|}+\xi^2_{\perp}q^2_{\perp}\right).
\label{EQN:Gamma_Q}
\end{align}

The solid lines in Fig.~\ref{FIG:Qcuts_with_AB_phase}c-h are fits to the $\mathbf{Q}$-dependence predicted by  Eqn.~\ref{EQN:MMP} for each energy transfer with the corresponding extracted widths, $\sigma_{\|}(\omega)$ and $\sigma_{\perp}(\omega)$, plotted in \ref{FIG:Qcuts_with_AB_phase}i.  The solid lines in \ref{FIG:Qcuts_with_AB_phase}i show the predictions the MMP form (Eqn.~\ref{EQN:MMP}) with the fitted parameters in Table.~\ref{table:fitted_parms}.  In addition, the $\omega$-dependence  of $\chi^{\prime\prime}(\mathbf{Q},\omega)$ at $\mathbf{Q}_\delta$ is well described by the overdamped response (Eq.~\ref{EQN:MMP}) at $B=7.95$\,T and other fields as shown in Fig.~\ref{FIG:Ecuts}a-f. Thus the MMP form provides an excellent description of our data.

The lowest energy and strongest fluctuations occur at $\mathbf{Q}_{\delta}$. These are parameterised by the relaxation rate $\Gamma_{\delta}$ and susceptibility $\chi_{\delta}$. The field dependence of these parameters is shown in Fig.~\ref{FIG:Ecuts}g,h. We see a dramatic softening (Fig.~\ref{FIG:Ecuts}h) and an increase in strength (Fig.~\ref{FIG:Ecuts}g) of the fluctuations on approaching $B_c$.  At $B=7.95$\,T, the fluctuating moment associated with the low-energy ($<$1~meV) excitations is $\sqrt{\langle m^2 \rangle} \approx 0.17(2)$\,$\mu_B$\,Ru$^{-1}$ which is larger than the ordered moment $\langle m \rangle$=0.1\,$\mu_B$\,Ru$^{-1}$ \cite{Lester2015_LRPC}.

We can also measure at other reciprocal space positions to probe the polarisation of the excitations. Fig.\;\ref{FIG:Ecuts}j shows data collected using the IN14 spectrometer at the Institut Laue-Langevin for $\hbar \omega=0.3$~meV and fields through $B_c$.  At the $(1-\delta,0,0)$ position we see that the field-dependent intensity mirrors the behaviour at $(\delta,0,0)$ in Fig.~\ref{FIG:Ecuts}g, with a peak in SDW phases near $B_c$. In contrast, no such peak is observed at $(1,\delta,0)$ implying that the spin fluctuation mode (SFM) is only seen when the propagation vector (reduced to the first Brillouin zone) is parallel to $\mathbf{Q}$. Using the standard theory for magnetic INS (See Methods for details.) we can infer that the spin fluctuations associated with the soft mode propagating along the $a$-axis are polarised along the $b$-axis (see Fig.~\ref{FIG:Ecuts}j). The polarisation is perpendicular to the $c$-axis and the propagation vector, and in the same direction as the ordered moment of the SDW \cite{Lester2015_LRPC}. Thus the soft excitation near $B_c$ is found to be a transverse spin fluctuation mode polarised within the RuO$_2$ planes.  

\begin{figure*}[t]
\begin{center}
\includegraphics[width=0.99\linewidth]{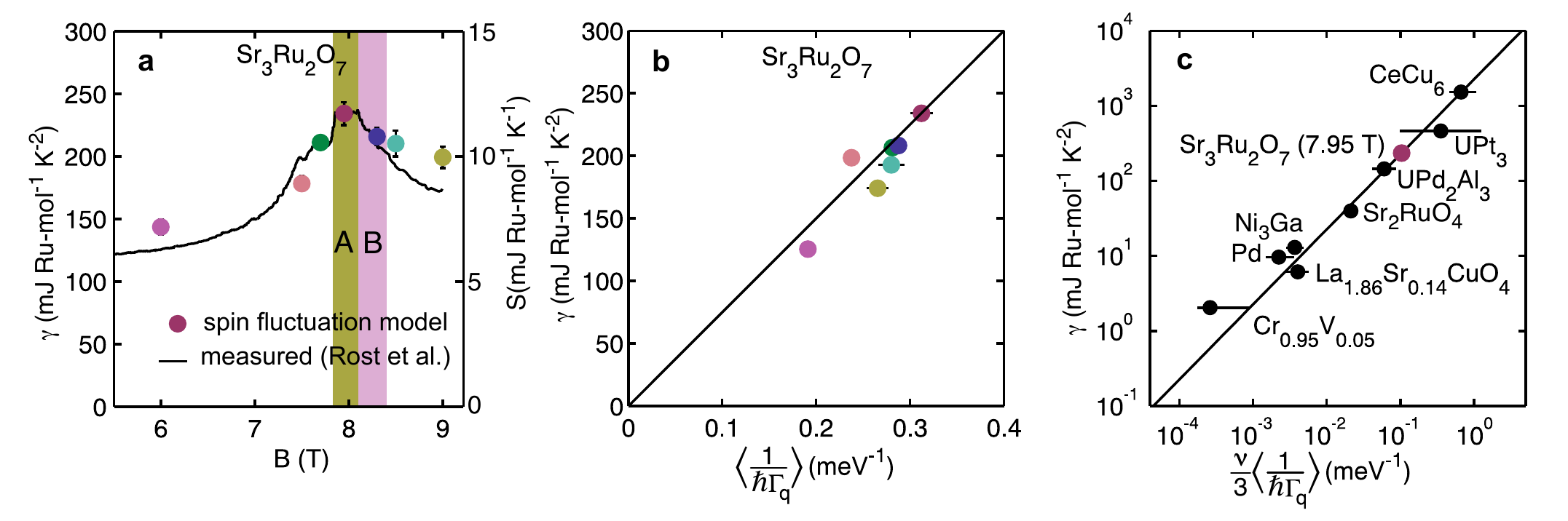}
\end{center}
\caption{\textbf{Spin fluctuations and the specific heat capacity.} \textbf{a} Circles show the low-temperature linear co-efficient of the heat capacity $\gamma$ calculated using the phenomenological spin-fluctuation model described in the text and our experimentally determined parameters. Solid line is the measured $\gamma$ of Rost et al. \cite{Rost2011_RGBP}. Note the spin-fluctuation model predicts increases of $\gamma$ approaching $B_c$ and on entering SDW phases A and B. \textbf{b} The measured $\gamma$ \cite{Rost2011_RGBP} for field values in panel a (fields denoted by colours) plotted against the Brillioun-zone-averaged inverse spin relaxation rate $\langle \Gamma(\mathbf{Q})^{-1} \rangle$.  \textbf{c} The measured $\gamma$ plotted against the experimentally determined $\langle \Gamma(\mathbf{Q})^{-1} \rangle$ for a variety of correlated electron metals~\cite{Hayden2000_HDAP,Steffens2019_SSKM}. Solid line is  Eq.~\ref{eqn:Gamma_ave} and $\nu=3$ for all systems except Sr$_3$Ru$_2$O$_7$ where $\nu=1$.}
\label{FIG:gamma_B}
\end{figure*}

\section*{Discussion}
We have identified an overdamped collective transverse spin-fluctuation mode which is uniquely controlled by magnetic field and becomes soft near $B=B_c$. In this section, we discuss the origin of the mode and its relationship to the SDW order and strange metal properties of Sr$_3$Ru$_2$O$_7$.
The characteristic wavevectors of emerging magnetic order and low-energy magnetic excitations of quantum materials can often be understood in terms of nesting features of the underlying Fermi surface. In the case of Sr$_3$Ru$_2$O$_7$, density functional theory (DFT) calculations \cite{Efremov2019_ESRC} have shown that exchange splitting induced by a large magnetic field causes a topological Lifshitz transition (LT) in the Fermi surface (FS) which creates new pockets of the $\gamma$-band. The four $\gamma$-pockets are centred on the $X$ point of the BZ and derive from a FS reconstruction due to the RuO$_6$ octahedra rotations present in the Sr$_3$Ru$_2$O$_7$ structure \cite{Shaked2000_SJCI}. The presence of the $\gamma$-pockets allows nesting for a certain range of fields with wavevectors similar to those of the SDW order and the soft-magnetic excitations reported here.  We believe the SFM is a paramagnon-like excitation \cite{Doniach1966_DE} characteristic of metals close to magnetic order. Such modes are understood in terms of the Hubbard model where single-particle spin-flip excitations involving electronic states near the Fermi energy are enhanced by the electron interactions on the Ru site.  

Our results show that the SFM becomes stronger in the SDW ordered state (see Fig.~\ref{FIG:Ecuts}). Thus the SDW state in Sr$_3$Ru$_2$O$_7$ forms together with a background of strong quantum and thermal magnetic fluctuations.  
An interesting feature of Sr$_3$Ru$_2$O$_7$ is that at finite constant $T$ the entropy of the SDW phases is larger than in the surrounding regions at lower and higher fields (see Fig.\, \ref{FIG:gamma_B}a) \cite{Rost2009_RPMM}.  
This may be understood in terms of the data in Fig.~\ref{FIG:Ecuts}. If $\chi^{\prime\prime}(\mathbf{Q},\omega)$ is large then the fluctuation-dissipation theorem implies that strong spin fluctuations are present.  The increase in $\chi^{\prime\prime}(\mathbf{Q},\omega)$ observed for fields near $B=B_c$ (see Fig.~\ref{FIG:Ecuts}) means that, at the small but finite temperature investigated ($T=50$\,mK), thermal fluctuations would be induced increasing the entropy of the system as observed experimentally (this is estimated explicitly below).   
Our data supports the notion that the SDW state in  Sr$_3$Ru$_2$O$_7$ is stabilised by fluctuations i.e. the free energy is lowered by the system sampling more configurations. The presence of fluctuations can be inferred from the variation of entropy with field and our scattering measurements. This ``order-by-disorder'' mechanism whereby fluctuations stabilise an ordered phase has been discussed by Onsager, Villain and others \cite{Green2018_GCK} in a wide variety of systems ranging from colloidal suspensions to insulating magnets. 

The quantum critical and strange metal behaviour of Sr$_3$Ru$_2$O$_7$ have been revealed by $B$-dependent transport and thermal measurements \cite{Grigera2001_GPSC,Rost2009_RPMM,Rost2011_RGBP,Sun2018_SRPM}. Clearly it is important to establish how this behaviour relates to the spin fluctuations reported here.  The strong spin fluctuations we observe near $B_c$ could provide scattering \cite{Mousatov2020_MBH} resulting in strange metal $T$-linear behaviour transport if they have the appropriate $T$-dependence. This will be studied in future work. The signature of quantum criticality in the thermal properties of Sr$_3$Ru$_2$O$_7$ is illustrated in Fig.~\ref{FIG:Qcuts_with_AB_phase}a and Fig.~\ref{FIG:gamma_B}a. At finite temperature, there is a divergence in field of the specific heat $\gamma=C/T \sim 1/|B-B_c|$ approaching the SDW phases.  We can use our data to make a quantitative estimate of the electronic specific heat and establish whether the signatures seen in thermal measurements are due to spin fluctuations.

\begin{figure}[tb]
\begin{center}
\includegraphics[width=0.9\linewidth]{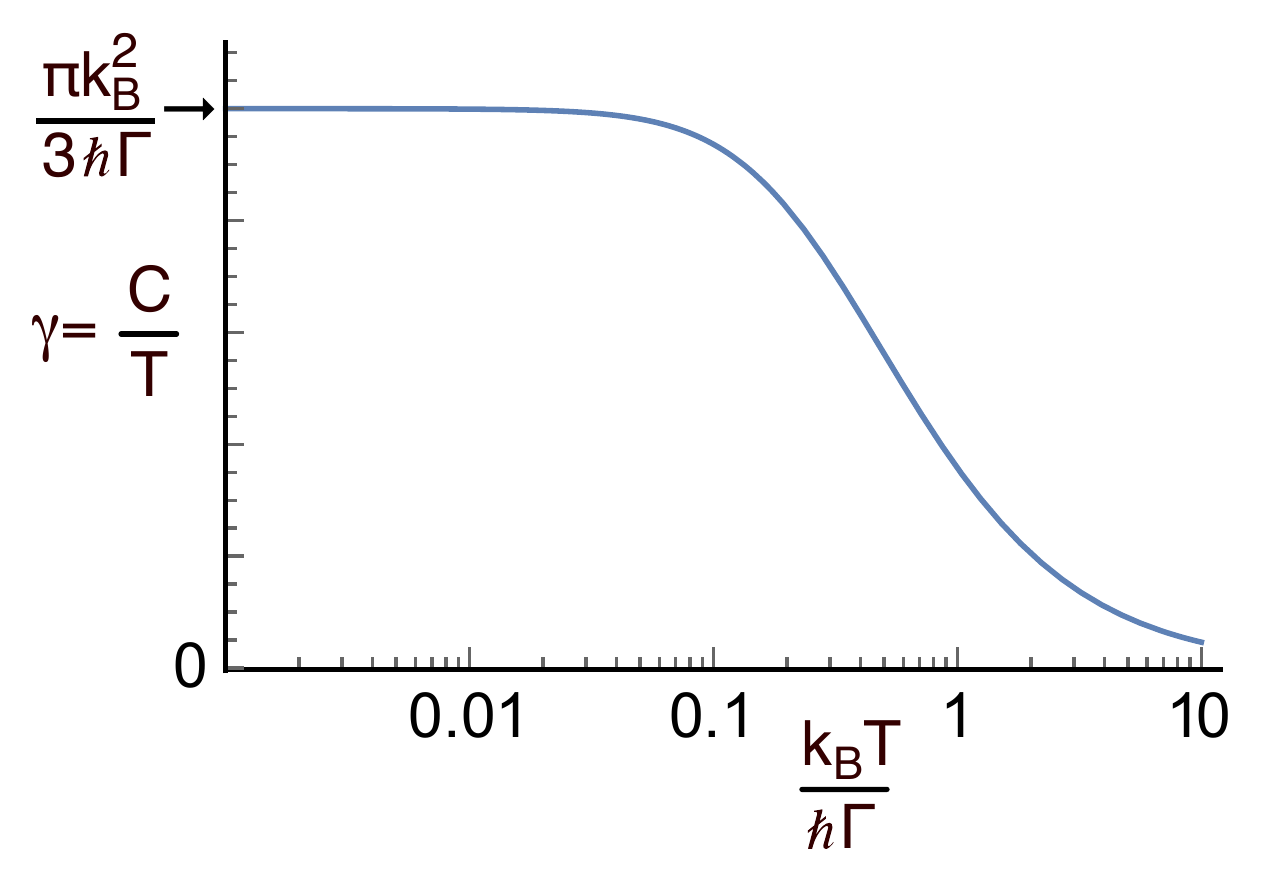}
\end{center}
\caption{The temperature dependence of $\gamma=C/T$ for a single spin-fluctuation mode with a temperature-independent relaxation rate $\Gamma$ calculated from Eq.~\ref{eqn:gamma}.}
\label{FIG:gammavsT}
\end{figure}

The low temperature thermal properties of metals near magnetic instabilities have been discussed in terms of spin fluctuation (paramagnon) theory originally developed in the 1960s \cite{Doniach1966_DE, Brinkman1968_BE} and ``self-consistent renormalization'' (SCR) or self-consistent one-loop approximation introduced in the 1970s \cite{Murata1972_MD,Moriya1985_T, Lonzarich1986_L,Edwards1992_EL,Moriya1995_MT,Ishigaki1999_IM}. These models address the enhancement of the low-temperature specific heat in metals that arises from the scattering of electrons by spin fluctuations.  The spin fluctuations concerned are strongly damped and can drastically alter the thermodynamic properties of the metals.
Edwards and Lonzarich \cite{Edwards1992_EL}, and Moriya and collaborators \cite{Moriya1995_MT,Ishigaki1999_IM} have obtained (See Methods for details) an approximate expression for the contribution of spin fluctuations to $\gamma$ in the $T \rightarrow 0$ limit. This is 
\begin{align}
\gamma = \frac{\nu \pi k_B^2}{3 \hbar} \left< \frac{1}{\Gamma(\mathbf{Q})} \right>_{\textrm{BZ}},  \label{eqn:Gamma_ave}
\end{align}
where $\nu$ is the number of polarisations of the spin fluctuations that contribute to the specific heat, $\Gamma(\mathbf{Q})$ is the spin-fluctuation relaxation rate, and $\langle \hdots \rangle$ denotes an average over the Brillouin zone (BZ).  Fig.~\ref{FIG:gammavsT} shows the heat capacity expected for a single mode with temperature independent $\Gamma(\mathbf{Q})$.  Eq.~\ref{eqn:Gamma_ave} may be checked experimentally by measuring $\Gamma(\mathbf{Q})$ throughout the Brillouin zone using inelastic neutron scattering. Fig.~\ref{FIG:gamma_B}c shows a comparison of the measured $\langle \Gamma(\mathbf{Q})^{-1} \rangle$ and heat capacity $\gamma$ for a variety of correlated metals(see Methods for details) demonstrating this method broadly works \cite{Hayden2000_HDAP,Steffens2019_SSKM}. We have assumed that $\nu=3$ for all the systems except Sr$_3$Ru$_2$O$_7$.

To estimate the low-$T$ specific heat of Sr$_3$Ru$_2$O$_7$ using Eq.~\ref{eqn:Gamma_ave} we set $\nu=1$ since only a single transverse spin fluctuation mode is observed (Fig.\,\ref{FIG:Ecuts}j). The magnetic response is parameterised using Eq.~\ref{EQN:MMP} with the parameters in Table.~\ref{table:fitted_parms} for $B=$7.95\;T. Our estimate (see Methods) yields $\gamma_{\textrm{SFM}} =0.23(3)$\;J\;K$^{-2}$\;Ru-mol$^{-1}$ which agrees with the measured value \cite{Rost2011_RGBP} $\gamma_{\textrm{exp}} =0.22$\;J\;K$^{-2}$\;Ru-mol$^{-1}$.

For $B=7.95$\;T we made INS measurements over a wide range of wavevectors and energies to demonstrate that Eq.~\ref{EQN:MMP} provides a good description of the magnetic response.  At other fields, data was only collected near the incommensurate position $\mathbf{Q}_{\delta}$ where the spin fluctuations are strongest and have the lowest energy scale. This allows the field dependence of $\chi_{\delta}$ and $\Gamma_{\delta}$, but not $\xi_{\|,\perp}$, to be determined (see Fig.\;\ref{FIG:Ecuts}g-h).  Near a critical point, we expect $\chi_{\delta}$ and $\Gamma_{\delta}$ to be controlled by the correlation length $\xi$. In mean-field theory (see Millis-Monien-Pines\cite{Millis1990_MMP} and Moriya and Ueda \cite{Moriya2000_MU}), we expect $\Gamma_{\delta} \propto \xi^{-2}$ and $\chi_{\delta} \propto \xi^2$. This is consistent with the data plotted in Fig.\,\ref{FIG:Ecuts}i where it is shown that $\tau=\Gamma_{\delta}^{-1} \propto \chi_{\delta}$. Thus, if we assume
\begin{align}
\hbar \Gamma_{\delta}(B) = c_{\|}/ \xi_{\|}^2(B) =  c_{\perp}/ \xi_{\perp}^2(B) 
\end{align}
and use the values of $c_{\|}$ and $c_{\perp}$ determined at  $B=7.95$\;T (Table.~\ref{table:fitted_parms}) we have a \textit{field-dependent} model for the magnetic response $\chi^{\prime\prime}(\mathbf{Q},\omega)$. This may be used to compute $\left< \Gamma^{-1}(\mathbf{Q}) \right>(B)$ and hence the field-dependence of the heat capacity.  The result of this procedure is shown by the points in Fig.~\ref{FIG:gamma_B}a. The heat capacity calculated from our spin fluctuation model shows a remarkable agreement with experimental measurements (solid line). The model reproduces the increase in $\gamma$ and entropy $S$ on approaching the SDW phases and also the increase of these quantities on entering the SDW phases. 
\begin{table}[tb]
\begin{center}
\begin{tabular}{|l|l|}
\hline
$\chi_{\delta}$ ($\mu_B^2$\;meV$^{-1}$f.u.$^{-1}$)  & 12.5(3) \\
$\delta$ (r.l.u) & 0.230(2) \\
$\xi_{\|}^{-1}$ (r.l.u.) &  0.0116(5)  \\
$\xi_{\perp}^{-1}$ (r.l.u.) & 0.022(1)   \\
$\hbar \Gamma_{\delta}$ (meV ) & 0.07(1)  \\ 
$c_{\|}$ (r.l.u$^2$) &  145(22) \\
$c_{\perp}$ ( r.l.u$^2$ ) & 522(86)   \\
\hline
\end{tabular}
\end{center}
\caption{Fitted susceptibility parameters for $B=$7.95\;T and $T=50$\;mK. $c_{\|}$ and $c_{\perp}$ are derived from other parameters.}
\label{table:fitted_parms}
\end{table}

We show how the collective spin fluctuations in the correlated electron metal Sr$_2$Ru$_3$O$_7$ evolve in a near singular way, as a magnetic field sweeps the system through a quantum critical point at very low temperature. These fluctuations may stabilise the SDW order observed in this system. We demonstrate how a simple phenomenological model describing the scattering of electrons by the collective spin fluctuations can compute the low temperature heat capacity and entropy of  Sr$_2$Ru$_3$O$_7$. Hence the novel thermal and transport properties of Sr$_2$Ru$_3$O$_7$ \cite{Grigera2001_GPSC,Grigera2004_GGBW,Rost2009_RPMM,Borzi2007_BGFP} can be understood in terms of the spin fluctuations reported here and the SDW order \cite{Lester2015_LRPC}. The model appears to have more general applicability to other materials with damped magnetic excitations. For example, it might be used to test whether the doping-dependent peak in the heat capacity $\gamma$ in cuprate superconductors \cite{Michon2019_MGBK} can be understood through magnetic excitations. 
 
\section{Methods}

\textbf{Sample growth and characterisation.} Our sample was an array of 9 single crystals sample with total mass of 6.6~g and was grown using an image furnace~\cite{Perry2004_PeMa}. The samples were co-aligned on thin aluminium plates with the [001] direction out of plane (parallel to applied field) and the sample had a total mosaic of 1.5$^\circ$. Susceptibility, transport and neutron diffraction measurements on samples used in the present experiment showed that they were of comparable purity to previous studies~\cite{Grigera2004_GGBW,Borzi2007_BGFP,Rost2009_RPMM}. In particular, they exhibit the hallmarks of the quantum critical phases shown in Figure~\ref{FIG:Qcuts_with_AB_phase}.

\textbf{Neutron scattering measurements.}  Experiments were performed at the LET spectrometer at the ISIS spallation source and the IN14 spectrometer at the ILL
at temperatures down to $T=50$ mK and fields up to $B=11$ T. For LET we used incident energies $E_i$=0.75, 1.4 and 3.5\,meV. INS can be used to probe the imaginary part of the generalised susceptibility $\chi^{\prime\prime}(\mathbf{Q},\omega)$, a measure of the strength of magnetic excitations at a particular $(\mathbf{Q},\omega)$. The magnetic scattering cross section is given by \cite{Boothroyd2020_A}
\begin{align}
\frac{d^2\sigma}{d\Omega dE}&=\frac{(\gamma_n r_e)^2}{\pi g^2\mu_B^2}\frac{k_f}{k_i}\frac{|\mathrm{F}(\mathbf{Q})|^2}{1-\exp(-\hbar\omega/kT)} \nonumber\\
&\times 
\sum_{\alpha\beta} \chi_{\alpha\beta}^{\prime\prime}(\mathbf{Q},\omega)(\delta_{\alpha\beta}-\hat{Q}_{\alpha}\hat{Q}_{\beta}),
\end{align}
where $(\gamma_n r_e)^2=0.2905$~barn sr$^{-1}$, $g$ is the Land\'e factor, $\mathbf{k}_i$ and $\mathbf{k}_f$ are the incident and final neutron wave vectors, $|\mathrm{F}(\mathbf{Q})|^2$ is the magnetic form factor for the Ru atom and $\mathbf{Q}=\mathbf{k}_i-\mathbf{k}_f$. Our data were placed on an absolute scale by comparing the scattering signal with that from a V standard.  Our quoted susceptibilities are $\chi^{\prime\prime}=\frac{1}{3}(\chi_{xx}^{\prime\prime}+\chi_{yy}^{\prime\prime}+\chi_{zz}^{\prime\prime})$ and the factor $(\delta_{\alpha\beta}-\hat{Q}_{\alpha}\hat{Q}_{\beta})$ is used to determine the polarization of the excitations. 

\textbf{Spin fluctuation heat capacity model.}
Spin fluctuation theory in the 
 ``self-consistent renormalization'' (SCR) or self-consistent one-loop approximation \cite{Murata1972_MD,Moriya1985_T, Lonzarich1986_L,Edwards1992_EL,Moriya1995_MT,Ishigaki1999_IM} can be used to estimate the low-temperature  free energy $F$ can be expressed \cite{Brinkman1968_BE,Lonzarich1986_L} as 
\begin{align}
    F = \sum_{\mathbf{\nu}, \mathbf{Q}} \int_{0}^{\omega_c} \; d\omega \frac{F_{\text{osc}}(\omega)}{\pi} \frac{\Gamma_{\mathbf{\nu}}(\mathbf{Q})}{\omega^2+\Gamma_{\nu}^2(\mathbf{Q})}, \label{eqn:free_energy}
\end{align}
where $F_{\text{osc}}(\omega)=\hbar \omega/2+ k_B T \ln[1-\exp(-\hbar \omega/ k_B T)]$ is the free energy of a harmonic oscillator with frequency $\omega$ and $\Gamma_{\mathbf{\nu}}(\mathbf{Q})$ is the relaxation rate of a spontaneous spin fluctuation of wavevector $\mathbf{Q}$ and polarization $\mathbf{\nu}$.  Eq.~\ref{eqn:free_energy} may be used to obtain an approximate expression \cite{Ishigaki1999_IM} for the linear coefficient of specific heat $\gamma$,
\begin{align}
\gamma &= \frac{C}{T} = -\frac{\partial^2 F}{\partial T^2}  \\
&= \sum_{\mathbf{\nu}, \mathbf{Q}} \; \int_{0}^{\omega_c} \; d\omega \; \frac{C_{\text{osc}}(\omega)}{T} \frac{1}{\pi} \frac{\Gamma_{\nu}(\mathbf{Q})}{\Gamma_{\nu}(\mathbf{Q})^2+\omega^2}. \label{eqn:gamma}
\end{align}
This is the sum of the specific heat $C_{\text{osc}}$ of harmonic oscillators with the frequency distribution of the spin-fluctuation spectrum, where
\begin{align}
C_{\text{osc}}(\omega) = \frac{\hbar^2 \omega^2}{k_B T^2}  \frac{e^{\hbar\omega/k_B T}}{(e^{\hbar \omega/k_B T}-1)^2}. \label{eqn:C_osc}
\end{align}
To illustrate the result, we evaluate Eq.~\ref{eqn:gamma} numerically for a single $(\nu,\mathbf{Q})$ mode with temperature independent  $\Gamma_{\mathbf{\nu}}(\mathbf{Q})=\Gamma$ and $\omega_c \rightarrow \infty$. The result is shown in Fig.~\ref{FIG:gammavsT}.

A renormalization group (RG) analysis by Millis \cite{Millis1993_M} obtained the quantum critical behaviour for various magnetic models. In the case of the 2D antiferromagnet, which is of interest here, both SCR \cite{Moriya1995_MT,Ishigaki1999_IM} and RG \cite{Millis1993_M} theory yield a logarithmic contribution to $\gamma=C/T \sim \ln(1/T)$ at low temperatures. 

Edwards and Lonzarich \cite{Edwards1992_EL}, and Moriya and collaborators \cite{Moriya1995_MT,Ishigaki1999_IM} have obtained an approximate expression (Eq.~\ref{eqn:Gamma_ave}) for $\gamma$ in the $T \rightarrow 0$ limit based on Eqns.~\ref{eqn:free_energy}-\ref{eqn:C_osc}. 

\textbf{Application of the spin-fluctuation theory of the specific heat to Sr$_3$Ru$_2$O$_7$}

We estimate the low-$T$ specific heat of Sr$_3$Ru$_2$O$_7$ using Eq.~\ref{eqn:Gamma_ave} treating the spin fluctuations as being 2D and assuming the only a single transverse spin fluctuation mode contributes to the specific heat i.e. $\nu=1$. The magnetic response is parameterised using Eq.~\ref{EQN:MMP} with the parameters in Table~\ref{table:fitted_parms} for $B=$7.95\;T and $T=50$\;mK. In order to estimate the heat capacity using Eq.~\ref{eqn:Gamma_ave}, we average $\Gamma^{-1}(\mathbf{Q})$ over the BZ. Fluctuations corresponding to the four $\mathbf{Q}_{\mathbf{\delta}}$ wavevectors are believed to pervade the whole sample so no domain averaging is required.  The average is dominated by the regions near the four $\mathbf{Q}_{\mathbf{\delta}}$ wavevectors, where there is little overlap of the different terms in the sum of Eq.~\ref{EQN:MMP}. Thus $\langle \Gamma^{-1}(\mathbf{Q}) \rangle$ can be computed by averaging $\Gamma^{-1}(\mathbf{Q})$ using Eq.~\ref{EQN:Gamma_Q}.  We find, for $B = 7.85$\;T, $\langle (\hbar \Gamma)^{-1}(\mathbf{Q}) \rangle = 0.312 \pm 0.012$\;meV$^{-1}$. Hence $\gamma=N_A  \pi k^2_B /3  \times 0.312 =  0.233$\;J\;K$^{-2}$\;Ru-mol$^{-1}$.

\section{Data availability}
All relevant data are available from the corresponding authors upon reasonable request.

\section{Acknowledgements}
We acknowledge informative discussions with P. Coleman, A. Chubukov, R. Evans, A. Green, G. G. Lonzarich, M. Zhu and A. P. MacKenzie. Our work was supported by the UK EPSRC (Grant Nos. EP/J015423/1 and EP/R011141/1).

\section{Author contributions}
C.L. and R.S.P. prepared the samples. C.L., S.M.H., S.R., T.P.C., M.L., R.B., T.G., A.H., A.W., E.M.F. made neutron scattering measurements.  C.L. and S.M.H. analyzed the data and wrote the initial manuscript. All authors contributed to the discussion and provided feedback on the manuscript.

\section{Competing interests}
The authors declare no competing interests.

%

\end{document}